\documentclass[
superscriptaddress,
amsmath,amssymb,
aps,
twocolumn
]{revtex4-1}
\usepackage[
colorlinks=true,
linkcolor=blue,
filecolor=blue,
citecolor=blue,
urlcolor=blue,
]{hyperref}

\usepackage{graphicx}
\usepackage{dcolumn}
\usepackage{bm}
\usepackage{amsmath}
\usepackage{braket}
\usepackage{booktabs}
\usepackage{comment}
\usepackage{lineno}
\usepackage[separate-uncertainty=true,multi-part-units=single]{siunitx}

\begin{document}

\preprint{APS/123-QED}

\title{{Time-domain programmable beam-splitter operations \\for an optical phase-sensitive non-Gaussian state}}

\author{Daichi Okuno}
\affiliation{
    Department of Applied Physics, School of Engineering, The University of Tokyo,
    7-3-1 Hongo, Bunkyo-ku, Tokyo 113-8656, Japan
}
\author{Takato Yoshida}
\affiliation{
    Department of Applied Physics, School of Engineering, The University of Tokyo,
    7-3-1 Hongo, Bunkyo-ku, Tokyo 113-8656, Japan
}
\author{Ryoko Arita}
\affiliation{
    Department of Applied Physics, School of Engineering, The University of Tokyo,
    7-3-1 Hongo, Bunkyo-ku, Tokyo 113-8656, Japan
}
\author{Takahiro Kashiwazaki}
\affiliation{
    NTT Device Technology Labs, NTT Corporation,
    3-1, Morinosato Wakamiya, Atsugi, Kanagawa 243-0198, Japan
}
\author{Takeshi Umeki}
\affiliation{
    NTT Device Technology Labs, NTT Corporation,
    3-1, Morinosato Wakamiya, Atsugi, Kanagawa 243-0198, Japan
}
\author{Shigehito Miki}
\affiliation{
    Advanced ICT Research Institute, National Institute of Information and Communications Technology, 
    588-2 Iwaoka, Nishi, Kobe 651-2492, Japan
}
\author{Fumihiro China}
\affiliation{
    Advanced ICT Research Institute, National Institute of Information and Communications Technology, 
    588-2 Iwaoka, Nishi, Kobe 651-2492, Japan
}
\author{Masahiro Yabuno}
\affiliation{
    Advanced ICT Research Institute, National Institute of Information and Communications Technology, 
    588-2 Iwaoka, Nishi, Kobe 651-2492, Japan
}
\author{Hirotaka Terai}
\affiliation{
    Advanced ICT Research Institute, National Institute of Information and Communications Technology, 
    588-2 Iwaoka, Nishi, Kobe 651-2492, Japan
}
\author{Shuntaro Takeda}
\email{takeda@ap.t.u-tokyo.ac.jp}
\affiliation{
    Department of Applied Physics, School of Engineering, The University of Tokyo,
    7-3-1 Hongo, Bunkyo-ku, Tokyo 113-8656, Japan
}
\begin{abstract}
    We present a loop-based optical processor enabling time-domain programmable beam-splitter (BS) operations for a phase-sensitive non-Gaussian state. The loop itself is of high quality, allowing for storage of a non-Gaussian state for up to seven round trips while preserving its Wigner negativity and phase coherence. We perform various BS operations on a non-Gaussian state and evaluate them as transformations of the state's waveforms. Our work integrates non-Gaussian states with time multiplexing, laying the foundation for large-scale universal quantum information processing.
\end{abstract}
\maketitle
\section{Introduction}
    Encoding quantum information into optical continuous-variable states multiplexed in the time domain is one of the most efficient approaches toward large-scale quantum information processing (QIP)~\cite{Takeda.2019-2}. In this scheme, an indefinitely large number of optical pulses, which carry the quantum information, are arranged in the time domain and interact with each other sequentially through optical delay lines and phase-controlled beam splitters (BSs). In such architecture, using on-demand multi-mode Gaussian states as inputs, there has been remarkable progress such as the generation of large-scale quantum entangled states~\cite{Larsen.2019, Asavanant.2019, Takeda.2019-1} and implementation of single-mode multi-step~\cite{Enomoto.2021, Asavanant.2021} or multi-mode multi-step~\cite{Larsen.2021, Yonezu.2023} quantum gates.
    
    In spite of these achievements, non-Gaussian quantum states have not yet been incorporated into time multiplexed multi-mode processing, though they are indispensable for achieving quantum advantage. It is well known that a circuit consisting solely of Gaussian input states, homodyne measurements, and linear operations such as BS operations and phase shifts can be efficiently simulated by a classical computer~\cite{Bartlett.2002}. One way to go beyond such a classically simulatable region is to prepare a multi-mode Gaussian state via a linear optical network and measure it in the Fock basis, an approach known as Gaussian boson sampling (GBS)~\cite{Hamilton.2017}, which is however applicable to only specific problems. Toward universal quantum computing architecture, we need the ability to treat non-Gaussian quantum states such as a cubic phase state or an ON state as a resource state for non-Gaussian gate operations~\cite{Marek.2011, Sabapathy.2018}, or a bosonic-qubit state for fault-tolerance~\cite{Cochrane.1999, Gottesman.2001}. The combination of such phase-sensitive non-Gaussian states and time multiplexed multi-mode processing has, however, been technically hindered by the randomness of the generation timing of non-Gaussian states with typical heralding methods. The processing circuit thus needs to be dynamically synchronized with the heralding signal. In previous experiments, such synchronization has been introduced only for the optical storage of phase-insensitive quantum states~\cite{Bouillard.2019, Cotte.2022}. These lack mechanisms for phase control and multi-mode operations, resulting in incompatibility with time-multiplexed continuous-variable QIP.
    \begin{figure}[b]
        \centering
        \includegraphics[width = 1\columnwidth]{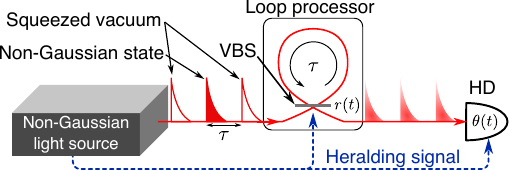}
        \caption{The conceptual illustration of our system. The amplitude reflectivity $r$ of the variable beam splitter (VBS) and the phase $\theta$ of the homodyne detector (HD) are controlled in synchronization with the non-Gaussian state heralding signal. \label{fig:concept}}
    \end{figure}
    \begin{figure*}[htbp]
        \centering
        \includegraphics[width = 1.0\linewidth]{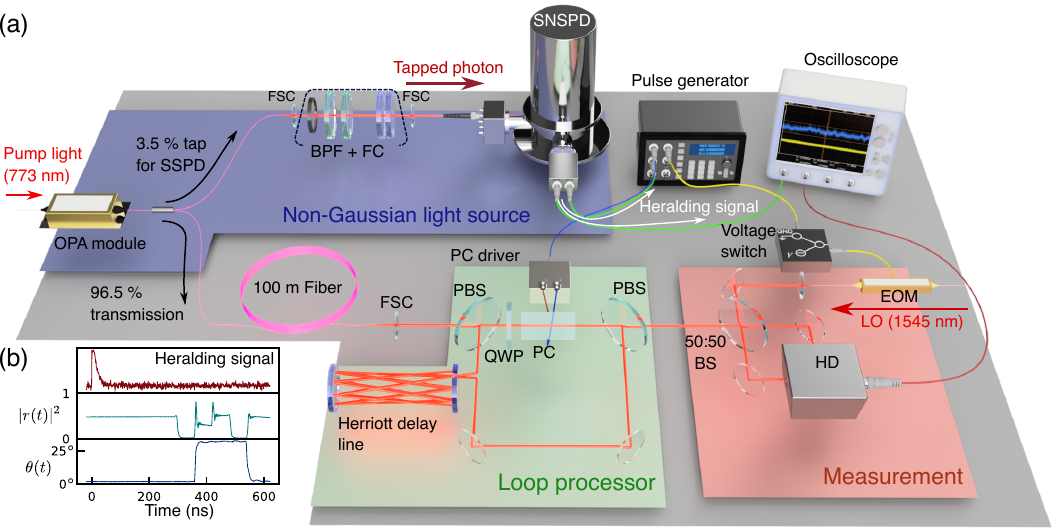}
        \caption{(a) The detailed schematic of the experimental system. BPF: Band-pass filter, BS: Beam splitter, EOM: Electro-optic modulator, FC: Filtering cavity, FSC: Free-space coupler, HD: Homodyne detector, LO: Local oscillator, OPA: Optical parametric amplifier, PBS: Polarizing beam splitter, PC: Pockels cell,  QWP: Quarter-wave plate, SNSPD: Superconducting nanostrip single-photon detector.
        (b) Typical time sequences of $r^2$, $\theta$ synchronized with the heralding SNSPD signal. \label{fig:setup}}
    \end{figure*}
    
    In this article, we present our loop-based optical processor which enables programmable multi-mode BS operations in the time domain, incorporating a probabilistically generated non-Gaussian state as an input (Fig.~\ref{fig:concept}). In principle, this system enables programmable BS operations on neighboring two modes, in a sequential and unbounded manner.
    For the characterization of the processor, we first demonstrate the storage and retrieval of a non-Gaussian state, maintaining the Wigner negativity for up to seven round trips of the loop with phase coherence. We then present the results of BS operations on a non-Gaussian wavepacket and its neighboring wavepackets. We show the transformation of a temporal mode function of a non-Gaussian state as proof of the proper BS operations.
    The developed system in this work already has useful applications in storage and waveform transformation of quantum states, and iterative protocols to generate exotic quantum states~\cite{Etesse.2014, Laghaout.2013, Weigand.2018}. Moreover, it is a key building block to combine non-Gaussian states with time-multiplexed QIP, paving the way for large-scale universal QIP based on not only loop-based quantum computing architecture~\cite{Takeda.2017, Su.2019} but also measurement-based one using a cluster state~\cite{Menicucci.2011, Takeda.2019-1}.
    
\section{Experimental setup}
    The conceptual structure of the developed system is shown in Fig.~\ref{fig:concept}. The entire system consists of three parts: the non-Gaussian light source, the loop processor, and the measurement part. The source outputs a non-Gaussian state only when photon detection heralds and otherwise it outputs a squeezed vacuum state. The real amplitude reflectivity $r$ of the variable beam splitter (VBS), as well as the phase $\theta$ of the local oscillator (LO) of the homodyne measurement, is controlled in synchronization with the heralding signal to perform the programmable BS operations on a non-Gaussian state and the subsequent characterization. Typical time sequences of $r^2$, $\theta$, and the heralding signal are shown in Fig.~\ref{fig:setup}(b).

    Figure~\ref{fig:setup}(a) is the detailed setup. Continuous-wave light from a fiber laser with a central wavelength of \SI{1545}{nm} is amplified by an Er-doped fiber amplifier and is used as a light source.
    Frequency-doubled 773-nm light by second-harmonic generation is fed to a fiber-coupled optical parametric amplifier (OPA) module as a pump for the generation of the squeezed vacuum $\hat{S}\ket{0}\equiv\ket{\mathrm{sq}}$, where $\hat{S}$ and $\ket{0}$ is the squeezing operator and the vacuum state, respectively.
    The OPA adopts a periodically-poled lithium niobate crystal in a waveguide. Owing to the cavity-free design of the module, the generated squeezed vacuum has a THz-order broad bandwidth~\cite{Kashiwazaki.2021, Kashiwazaki.2023}.
    For all experiments in this article, the squeezing level is set to \SI{2.6}{dB}, which corresponds to the pump power of $\sim$\SI{20}{mW}.
    
    A non-Gaussian state is obtained by the photon-subtraction technique~\cite{Wakui.2007,Neergaard-Nielsen.2006}. A small fraction \SI{3.5}{\percent} of the squeezed vacuum is tapped by a fiber beam splitter to be sent to a superconducting nanostrip single-photon detector (SNSPD)~\cite{Miki.2017}. Since the bandwidth of the squeezed vacuum significantly exceeds that of our homodyne detector ($\sim \SI{200}{MHz}$), a dielectric band-pass filter and two filtering Fabry-Perot cavities are placed before the SNSPD to set the bandwidth of the heralded state within that of the detector. The SNSPD click heralds a squeezed single photon state $\hat{a}\hat{S}\ket{0} \propto \hat{S}\hat{a}^\dagger\ket{0}\equiv\ket{\mathrm{cat}}$~\cite{Yoshikawa.2017}, an approximate state of a Schr\"{o}dinger's cat state, in the transmitted line of the fiber beam splitter. Here $\hat{a}$ denotes the photon annihilation operator. In our work, photon detection is performed with an overall efficiency of $\sim$\SI{20}{\percent}.
    
    \begin{figure*}[htbp]
        \centering
        \includegraphics[width = 1.0\linewidth]{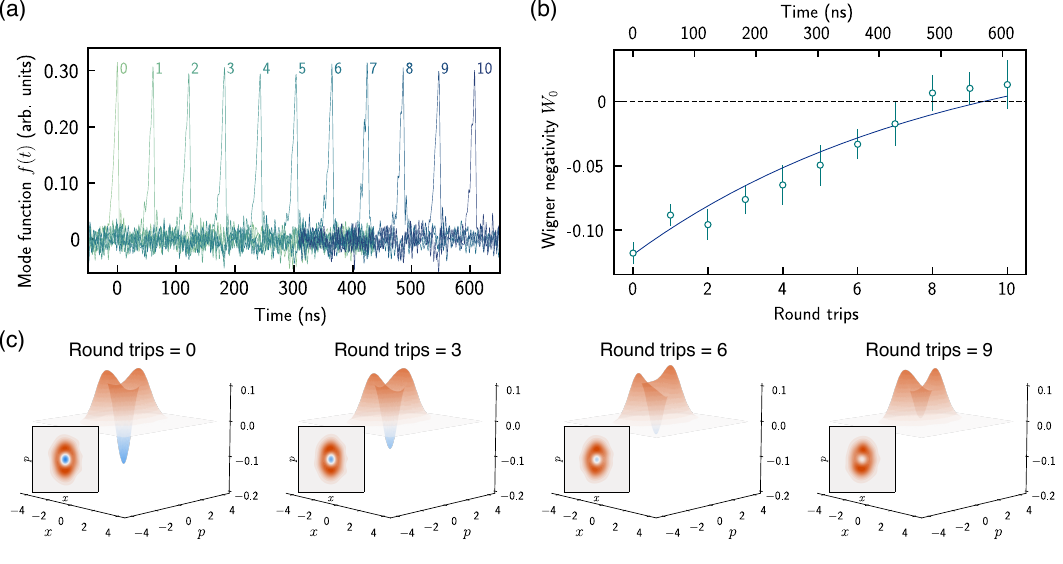}
        \caption{
            Performance of the loop-based processor as a quantum memory.
                (a) Obtained temporal mode functions for round trips $N=0$ to 10. Zero at the horizontal axis is set to the peak of the mode function for $N=0$.
                (b) The retention of the Wigner negativity at the origin of the phase space. The solid line shows the fitted curve using the loss inside ($\kappa_{\mathrm{in}}$) and outside of the loop ($\kappa_{\mathrm{out}}$) as fitting parameters. The obtained fitted parameters are $\kappa_{\mathrm{in}} = \SI{3.9 \pm 0.2}{\percent}$ and $\kappa_{\mathrm{out}} = \SI{28.3\pm 0.7}{\percent}$.
                (c) The Wigner functions for $N=0,3,6,9$. $x$ and $p$ are in-phase and out-of-phase quadrature of the light, respectively.
                The normalization convention $\hbar = 1$ is employed for the calculation of the Wigner function.
                \label{fig:memory}
        }
    \end{figure*}
    The basic constitution of the loop processor is the same as that of our previous experiments~\cite{Takeda.2019-1, Enomoto.2021, Yonezu.2023}: two polarizing beam splitters (PBSs) and a Pockels cell functioning as a VBS, and a Herriott-type optical delay line for the loop. The length of the loop is \SI{18.2}{m}, corresponding to a cycle time of $\tau = \SI{60.8}{ns}$.
    Synchronized with the SNSPD signal, a pulse generator sends pulse signals to the Pockels cell driver and a voltage switch for a fiber electro-optic modulator, enabling the dynamical control of $r$ and $\theta$. To take account of the delay of the electronics, a \SI{100}{m} fiber optical delay line is inserted between the non-Gaussian light source and the loop processor. The electronic delay is around \SI{270}{ns}, out of which $\sim \SI{130}{ns}$ comes from the pulse generator, $\sim \SI{60}{ns}$ from the Pockels cell driver, and $\sim \SI{80}{ns}$ from the electronic cables, respectively.

    After passing through the loop processor, the output state is combined with the LO and fed into the homodyne detector. To reconstruct the Wigner function, 3,000 data points are acquired for each LO's relative phase spanning from \SI{-90}{\degree} to \SI{75}{\degree} at \SI{15}{\degree} interval. The voltage switch changes $\theta$ from zero to a preset value at the measurement timing.
    Triggered by the SNSPD signal, the output of the homodyne detector is acquired by an oscilloscope, and the quadrature amplitude is extracted from the raw data by numerically applying the mode function filter. Finally the Wigner function is reconstructed by the maximum-likelihood technique~\cite{Lvovsky.2004} without loss correction. 
    
    The temporal mode function $f(t)$ of the heralded state can be obtained as a time-reversal of the temporal response function $h(t)$ of the frequency filter of the heralding photon: $f(t)\propto h(-t)$~\cite{Yoshikawa.2017}.
    Because the temporal response function of a cavity with a HWHM of $\gamma$ is given by $\mathrm{e}^{-\gamma t}\Theta(t)$, the temporal response function of the overall filters which is given as the convolution of them becomes
    \begin{align}
        h(t) &\propto
        \int_{-\infty}^{\infty} \mathrm{d}t^\prime \mathrm{e}^{-\gamma_1 (t-t^\prime)-\gamma_2 t^\prime}\Theta(t-t^\prime)\Theta\left(t^\prime\right)\\
        &\propto
        \left(\mathrm{e}^{-\gamma_1 t} - \mathrm{e}^{-\gamma_2 t} \right)\Theta(t). \label{eq:modefunc}
    \end{align}
    Here $\gamma_1$ and $\gamma_2$ are the cavities' bandwidth (HWHM) and $\Theta$ is the Heaviside step function. Note that the contribution from the band-pass filter is negligible.
    
    We use a coherent beam as a reference to get error signals for stabilizing the loop length, the length of the filtering cavities, the parametric gain of the OPA, and the relative phase of the LO. 
    As for the filtering cavity stabilization, the beam propagates in a reverse direction so as not to shine the SNSPD, while for all other stabilization it propagates in the same direction as the OPA output.
    During the measurement, the reference beam is blocked using two acousto-optic modulators, and the voltage applied to each piezo actuator is held constant (\textit{Sample\&Hold}~\cite{Wakui.2007}). In addition to it, during the stabilization period, two solid-state optical switches are used to prevent the reference light from entering into the SNSPD. The click count per second with the reference blocked is $\sim$36,000 while $\sim$500 out of them are fake counts resulting from stray light.
    
\section{Results}
    We first demonstrate the ability to store and retrieve a non-Gaussian state to characterize the quality of the loop. 
    For this operation, $r$ is changed according to the following steps: triggered by the SNSPD detection signal, firstly, the VBS is set to be transparent to let the state in the loop and then changed to maximally reflecting for the desired cycle times until it is again made transparent to retrieve the state.
    The successful storage of the state for $N$ cycle times results in temporal translation of the mode function by $N\tau$. 
    This is confirmed by assessing the mode functions of the stored states. Figure~\ref{fig:memory}(a) shows the experimentally obtained mode functions derived by the eigenfunction expansion of the autocorrelation function of the homodyne output~\cite{Morin.2013}. It indicates the expected behavior of the temporal mode function. 
    \begin{figure*}[htbp]
        \centering
        \includegraphics[width = 1.0\linewidth]{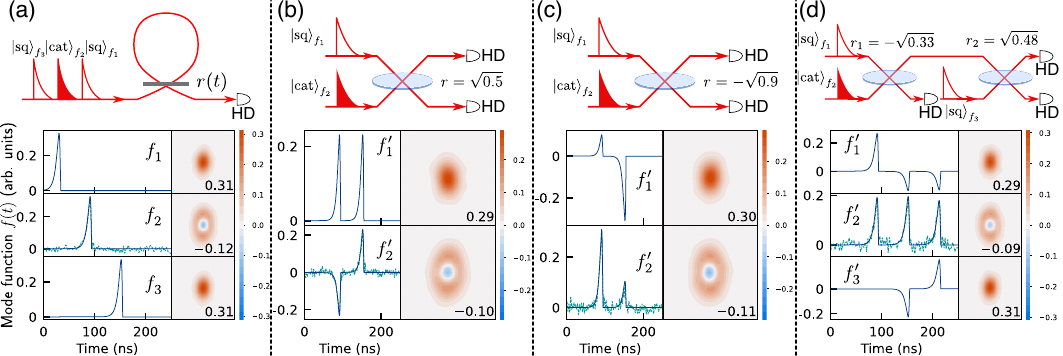}
        \caption{
            Temporal mode functions and the corresponding Wigner functions ($\hbar = 1$) of the input (a) and the output (b)-(d) of the multi-mode BS operations on a squeezed single photon state and squeezed vacuum states.
            The top row shows the sketch of the temporal relation of the input (a) and the equivalent optical circuits in the path encoding (b)-(d). HD stands for homodyne detector.
            The solid line in the mode function panel is the obtained temporal mode function from Eqs.~\eqref{eq:modefunc} and \eqref{eq:reshape}, and the dashed line is the one calculated from the eigenfunction expansion. The horizontal and the vertical direction of the Wigner function panel correspond to $x$ and $p$ in Fig.~\ref{fig:memory}, respectively. $W_0$ for each state is written in the panel.
            \label{fig:reshape}
        }
    \end{figure*}
 
    To further characterize the performance of the loop as a memory, the value of the Wigner function at the origin of the phase space, denoted as $W_0$, is evaluated. For calculation of the quadrature amplitude, we use the mode function $f(t-t_0-N\tau)$ for $N$ round trips data, where $t_0$ is the click timing. $t_0, \gamma_1$, and $\gamma_2$ are determined from the data for $N=0$ such that they maximize the variance of the quadrature amplitudes. The obtained values for $\gamma_1$ and $\gamma_2$ are $2\pi\times\SI{28.1}{MHz}$ and $2\pi\times\SI{100}{MHz}$, respectively.
    Figure \ref{fig:memory}(b) shows the dependency of $W_0$ on the round trip number of the loop. The error bars are evaluated from the Fisher information at the estimated density matrix~\cite{Rehacek.2008}. As can be seen from the figure, the memory process retains the negativity of the state for up to seven round trips or \SI{426}{ns}. The decay of the negativity is analyzed by fitting, assuming the deterioration stems solely from the loss inside ($\kappa_{\mathrm{in}}$) and outside ($\kappa_\mathrm{out}$) the loop. The fitted values for $\kappa_{\mathrm{in}}$ and $\kappa_{\mathrm{out}}$ are \SI{3.9 \pm 0.2}{\percent} and \SI{28.3\pm 0.7}{\percent}, respectively. This corresponds to a lifetime of $\tau/\kappa_\mathrm{in} = \SI{1.6\pm0.08}{\micro s}$.
    We also remark that the phase information is saved as well as the negativity as shown in Fig.~\ref{fig:memory}(c).
    The achieved lifetime of the memory is comparable to or better than those achieved in previous experiments of loop-based storage of phase-insensitive states~\cite{Bouillard.2019} or cavity-based storage of phase-sensitive states~\cite{Hashimoto.2019}.

    We then demonstrate the BS operation on a non-Gaussian state and squeezed vacuum states in adjacent temporal modes programmably (Fig.~\ref{fig:reshape}(a)). The BS operator $\hat{\mathrm{BS}}(r)$ between temporal modes $f_1$ and $f_2$ with the amplitude reflectivity $r$ is characterized by the following input-output relation of the annihilation operators:
    \begin{align}
        \hat{\mathrm{BS}}(r)
        \begin{pmatrix}
            \hat{a}_{f_1}\\\hat{a}_{f_2}
        \end{pmatrix}
        \left(\hat{\mathrm{BS}}(r)\right)^\dagger
        =
        \begin{pmatrix}
            \sqrt{1-r^2} & r\\
            -r & \sqrt{1-r^2}
        \end{pmatrix}
        \begin{pmatrix}
            \hat{a}_{f_1}\\
            \hat{a}_{f_2}
        \end{pmatrix}.\label{eq:a_io}
    \end{align}
    $\hat{a}_f$ is the annihilation operator of the temporal mode function $f(t)$ which is defined as 
    \begin{equation}
        \hat{a}_f = \int \mathrm{d}t f(t)\hat{a}(t), \label{eq:af}
    \end{equation}
    using $\hat{a}(t)$, the annihilation operator at time $t$.
    From Eq.~\eqref{eq:a_io}, one can immediately find the BS operation on a squeezed vacuum in a temporal mode $f_1$ and a squeezed single photon in $f_2$ with the same squeezing parameter ends up in an entangled state:
    \begin{align}
        \ket{\psi} &= \hat{\mathrm{BS}}(r)\ket{\mathrm{sq}}_{f_1}\ket{\mathrm{cat}}_{f_2}\label{eq:psi}\\
        &=-r\ket{\mathrm{cat}}_{f_1}\ket{\mathrm{sq}}_{f_2} + \sqrt{1-r^2}\ket{\mathrm{sq}}_{f_1}\ket{\mathrm{cat}}_{f_2}.\label{eq:psi_entangle}
    \end{align}
    $\ket{\mathrm{sq}}_{f}$ and $\ket{\mathrm{cat}}_{f}$ are the squeezed vacuum and the squeezed single photon state in the temporal mode given by a mode function $f(t)$. From Eq.~\eqref{eq:psi} to Eq.~\eqref{eq:psi_entangle}, we used an identity that a product state of the same squeezed vacuum doesn't change under the beam-splitter transformation: $\hat{\mathrm{BS}}(r)\ket{\mathrm{sq}}_{f_1}\ket{\mathrm{sq}}_{f_2}  = \ket{\mathrm{sq}}_{f_1}\ket{\mathrm{sq}}_{f_2}$.
    
    On the other hand, from Eqs.~\eqref{eq:a_io} and \eqref{eq:af}, we can view this process as the transformation of the mode functions of the input states~\cite{Takase.2019}. Namely,
    \begin{equation}
        \ket{\psi} = \ket{\mathrm{sq}}_{f_1^\prime}\ket{\mathrm{cat}}_{f_2^\prime}, \label{eq:mode-trans}
    \end{equation}
    where $f_1^\prime(t)$ and $f_2^\prime(t)$ are the linearly transformed mode functions of the $f_1(t)$ and $f_2(t)$ given by
    \begin{align}
        \begin{split}        
            f_1^\prime(t) &= \sqrt{1-r^2} f_1(t) + r f_2(t),\\
            f_2^\prime(t) &= -r f_1(t) + \sqrt{1-r^2} f_2(t).
        \end{split}
        \label{eq:reshape}
    \end{align}
    
    From such a viewpoint, this operation can also be interpreted as an implementation of direct and \textit{in-situ} reshaping of a waveform of a non-Gaussian state, different from a conventional waveform manipulation technique by the frequency filtering of the heralding photon~\cite{Takeda.2013-nature,Takeda.2013-PRA,Yoshikawa.2018,Takase.2022}. The same discussion holds for the operation involving three or more modes. For the sake of experimental convenience of the characterization of the output, we adopt this interpretation and evaluate the BS operation by how it transforms the temporal mode functions of input states.

    We present the experimental results of the BS operations on two or three modes in Fig.~\ref{fig:reshape}.
    We adopted the following three experimental conditions for this demonstration as illustrated in the top row of Fig.~\ref{fig:reshape}(b)-(d): a two-mode BS operation with $r = \sqrt{0.5}$ (b), $r=-\sqrt{0.9}$ (c) and a three-mode BS operation with two beam splitters with $r_1 = -\sqrt{0.33}$ and $r_2 = \sqrt{0.48}$ (d). 
    The mode functions and the Wigner functions of input states are shown in Fig.~\ref{fig:reshape}(a). 
    The reshaped mode functions by the BS operation given by Eq.~\eqref{eq:reshape} are plotted at the bottom of Fig.~\ref{fig:reshape}(b)-(d). 
    The mode functions of the squeezed single photon states, obtained via eigenfunction expansion~\cite{Morin.2013}, are overplotted for reference. A good agreement between them indicates the waveform transformation of the non-Gaussian state is properly executed. It is worth noting that the mode function shown in Fig.~\ref{fig:reshape}(b) is particularly suitable for continuous-variable experiments owing to its dual peaks with opposite phases, which cancel out the DC component of the homodyne output~\cite{Asavanant.2019, Enomoto.2021, Yonezu.2023, Asavanant.2021}.
    
    To verify the mode is transformed without affecting the quantum state itself, we extract the quadrature amplitude using these mode functions. The reconstructed Wigner function is plotted alongside each temporal mode function. We observe only a slight degradation of $W_0$ of the non-Gaussian state for each transformed mode, while the Wigner functions of the squeezed vacuum states are kept almost the same. These results ensure the success of the BS operation on a non-Gaussian state and squeezed vacuum states in the time domain.
    
\section{Discussion}
    In conclusion, we have developed an optical loop-based and dynamically controllable processor synchronized with heralding signals of photon subtraction. We have successfully stored and retrieved the generated phase-sensitive non-Gaussian state, followed by its characterization by the homodyne measurement. The Wigner negativity is confirmed to preserve for up to seven round trips in the loop. We have further proved the processor properly performs BS operations on a non-Gaussian state and other states in neighboring temporal modes, which is equivalent to the \textit{in-situ} transformation of a non-Gaussian state waveform. This work is a fundamental achievement to scale up continuous-variable QIP in the time domain. The processor can be straightforwardly upgraded to a universal one by incorporating a phase shifter and a feedforward system inside the loop, in addition to an outer loop for memory~\cite{Takeda.2017}. Moreover, by leveraging recent advancements toward the generation of highly non-classical optical states such as cubic phase states~\cite{Yukawa.2013} and Gottesman-Kitaev-Preskill states~\cite{Konno.2024}, our system opens up avenues for applying optical QIP to the challenging practical tasks.

    \begin{acknowledgments}
        This work was partly supported by JST Grant Numbers JPMJFR223R, JPMJMS2064, and JPMJPF2221, JSPS KAKENHI Grant Numbers 23H01102 and 23K17300, the Canon Foundation, and MEXT Leading Initiative for Excellent Young Researchers.
    \end{acknowledgments}
%

\end{document}